\documentclass[prd,preprint,amsmath,nofootinbib,superscriptaddress]{revtex4-1}
\usepackage{bm}
\usepackage{slashed}
\usepackage{epsfig}
\usepackage{amsmath,latexsym}

\usepackage{graphicx}
\usepackage{epsfig}
\usepackage{hyperref}
\usepackage[usenames,dvipsnames]{color}
\newif\ifpdf
\ifx\pdfoutput\undefined
\pdffalse % we are not running PDFLaTeX
\else
\pdfoutput=1 % we are running PDFLaTeX
\pdftrue
\fi

\def\OMIT#1{}

\newcommand{\bea}{\begin{eqnarray}}
\newcommand{\eea}{\end{eqnarray}}

\newcommand{\beq}{\begin{equation}}
\newcommand{\eeq}{\end{equation}}

\begin{document}
%%%%%%%%%%%%%%%%%%%%%%%%%%%%%%%%%%%%%%%%%%
%Some more stuff to get graphics to work
\ifpdf
\DeclareGraphicsExtensions{.pdf, .jpg}
\else
\DeclareGraphicsExtensions{.eps, .jpg}
\fi
%%%%%%%%%%%%%%%%%%%%%%%%%%%%%%%%%%%%%%%%%%

\title{Gravitational spin Hamiltonians from the S matrix}

\author{Varun Vaidya}
\affiliation{Department of Physics, Carnegie Mellon University\footnote{email:vpvaidya@andrew.cmu.edu},
    Pittsburgh, PA 15213}

%%%%%%%%%%%%%%%%%%%%%%%%%%%%%%%%%%%%%%%%%%

\begin{abstract}
We utilize generalized unitarity and recursion relations combined with effective field theory(EFT) techniques to compute spin dependent interaction terms for inspiralling binary systems in the post newtonian(PN) approximation. Using these methods offers great computational advantage over traditional techniques involving Feynman diagrams, especially at higher orders in the PN expansion. As a specific example, we consider a binary system with one of the stars having non zero spin and reproduce the spin-orbit interaction up to 2.5 PN order as also the leading order $S^2$(2PN) Hamiltonian for an arbitrary massive object. We also obtain the $S^3$(3.5PN) spin Hamiltonian for an arbitrary massive object, which was till now known only for a black hole. Furthermore, we derive the missing $S^4$ Hamiltonian at leading order(4PN) for an arbitrary massive object and establish that a minimal coupling of a massive elementary particle to gravity  leads to a black hole structure. Finally, the Kerr metric is obtained as a series in $G_N$ by comparing the action of a test particle in the vicinity of a spinning black hole to the derived potential.
%We use generalized unitarity and recursion relations to obtain spin orbit interaction term for a compact binary system upto 2.5 PN order. This offers a demonstration of the huge computational advantage over traditional methods using feynman diagrams. We also obtain the 
\end{abstract}

\maketitle

\section{Introduction}
To observe gravitational waves, one needs very sensitive detectors due to to the tiny cross section of the waves with matter. There are several ground based detectors like VIRGO and LIGO (\cite{Aasi:2012rja},\cite{Abadie:2010cf}) which have a good chance of detecting gravitational waves in the next few years. For data analysis of such a signal, if and when it is discovered, it is necessary to have a theoretical template of the signal that is expected from inspiralling binary sources. While it is not possible to get an exact analytic solution for such a system in all regimes of it evolution, we can use approximate methods to get highly accurate analytic results especially in the slow motion and wide separation phase. The Hamiltonians for spinning and non spinning objects in the post newtonian approximation known to date are neatly listed in \cite{Hergt:2012zx}. These interactions have been derived using different formalisms, two of these being the ADM (\cite{Damour:2007nc},\cite{Steinhoff:2008zr},\cite{Steinhoff:2007mb},\cite{Hergt:2010pa},\cite{Levi:2011eq}) which compute the Hamiltonians and  NRGR (\cite{Goldberger:2004jt},\cite{Porto:2010tr},\cite{Porto:2008jj},\cite{Porto:2008tb}) which obtain the result in the form of a Lagrangian.\\
\\In this paper we extend the method introduced in \cite{Neill:2013wsa} to spinning sources, via effective field theory techniques using recent advances in S matrix calculations in particle physics. A similar approach was used in a recent paper to compute quantum gravity effects\cite{Bjerrum-Bohr:2013bxa}. We forego Einsteins point of view of treating gravity as a manifestation of space-time geometry and instead treat all effects of gravity as the propagation of a massless spin 2 particle on a flat background. Classical spinning objects are treated as local sources of gravitons and the modes which give rise to the classical potential between such objects are factorized from the radiative modes in an Effective Field Theory(EFT) (See \cite{Rothstein:2003mp} for review). For example, the technique of NRGR relies on explicit separation of scales relevant to the problem : the size of the objects $r_s$, the size of the orbit r and natural radiation wavelength r/v. Here the relative velocity v $<<$ c. Finite sized effects are treated by including new terms in the wordline action which are needed to regularize the theory. This usually involves terms obeying the correct symmetry constructed using the Riemann tensor and the velocity v. The accuracy in the PN expansion can be improved by adding higher dimensional operators. The coefficients of these operators are obtained by matching onto the full underlying theory which is GR. While doing calculations in such an EFT, Feynman diagrams will show up at the tree and loop level as perturbative techniques to iteratively solve for the Green's function of the full theory. \\
 Modern methods of computation for scattering amplitudes have dramatically reduced the effort involved in calculating loop amplitudes. Most of these involve the recursive use of on-shell amplitudes, which means that only the on-shell propagating modes of a field are used in any calculation. This technique automatically gets rid of the need for a gauge choice, thus eliminating the huge amount of redundancy involved in traditional Feynman  diagrams.\\ 
The most useful of these for our purposes is the BCFW recursion relation\cite{Britto:2004ap} and generalized unitarity methods(\cite{Bern:1994zx},\cite{Bern:1994cg}). These methods are traditionally applied for calculating on shell S matrix elements, but we are primarily interested in calculating the off shell potential between two spinning classical objects. The scattering amplitude is matched onto an effective theory in which the graviton is essentially integrated out. This leads to a well defined and IR finite classical potential. The calculated potential is a series in the relative velocity for a virialized orbit $v^2 \sim \frac{Gm}{r}$ and the spin. Both quantities count as 1PN in the post newtonian power counting parameter.\\
 This method has been applied for non spinning objects in \cite{Neill:2013wsa}. We extend this to the case of a binary system with one spinning components and demonstrate the use of this technique for calculating the spin orbit Hamiltonian to 2.5 PN order. We also present Hamiltonians for $S^2$, $S^3$ and $S^4$ terms at leading order for an arbitrary spinning object and show that a minimum coupling to gravity gives the interaction terms for a black hole. The Kerr metric is then derived as a series in the PN power counting parameter by expanding out the action of a test particle moving around a spinning black hole.

\section{Spin dependent Hamiltonians for compact binary systems}
In the calculations that follow, we obtain spin dependent Hamiltoninans from on-shell scattering amplitudes of a massive scalar particle with other massive particles with non-zero spin. From the addition rules of angular momentum, it is clear that scattering of a scalar with a particle of spin j will generate terms of 2j, 2j-1,...0 power in spin when we match onto an effective theory. For example, the scattering with a spin 1/2 particle will produce terms which are spin independent and linear in spin. Since we are interested in terms up to the $4^{th}$ order in the PN expansion, we need to go upto S=2, to generate the $S^4$ piece. While its true that all the relevant pieces that we need can be obtained by considering only the scattering with a $S=2$ particle, in order to obtain terms that are higher order in G, it is computationally efficient to consider the scattering of the smallest spin particle that can give us the required result. To that end we first consider the scattering of the scalar with a massive spin 1/2 particle to generate the spin-orbit Hamiltonian upto 2.5 PN order.\\   
For all the amplitudes that we calculate, we will need the three point interaction term of the scalar particle with a graviton. Assuming a minimal coupling to gravity gives us 
\begin{eqnarray}
M(p_3,p_4,m_b) = \frac{\kappa}{2}[p_{3\nu}p_{4\mu}+p_{3\mu}p_{4\nu}-\eta_{\mu \nu}(p_3 \cdot p_4 + m_b^2)]
\end{eqnarray}
where $p_3$ and $p_4$ are incoming momenta of the scalars with mass $m_b$, $\kappa = \sqrt{32\pi G_N}$.\\
We can also add a gauge invariant operator $R\phi^2$, but this does not affect the classical result.
For calculating loop amplitude, we will also need the on-shell 3 point amplitude in the spinor-helicity formalism (for a review see \cite{Dixon:1996wi},\cite{Peskin:2011in}).
where we use 3 and 4 in place of $p_3$ and $p_4$ respectively, while using  spinor-helicity notation.
\begin{equation}
iM(3,4,5^{+})= \frac{\kappa}{2}\frac{\langle r\slashed{3}5\rbrack^2}{\langle r5 \rangle^{2}}
\end{equation}
Here, r is any lightlike vector not proportional to the positive helicity graviton momentum 5. The amplitude for the negative helicity graviton is obtained by interchanging the angle and square brackets.\\
 For future use, we give the 4 point scalar graviton amplitude constructed using the BCFW recursion relation. This involves complexifying the momentum of two external massless particles while still maintaining momentum conservation. To apply this method, in principle, we need the theory to be BCFW constructible. This requires that the amplitude which is now a function of the complex variable z, should satisfy the condition  $\lim_{z \to \infty}M(z)/z = 0$. However, in our case this condition can be relaxed, since the terms that are not captured by the recursion do not contribute to the classical potential.  Also, we only need the 4 point amplitude with opposite helicities for the gravitons\cite{Neill:2013wsa}.

\begin{eqnarray}
M(3,4,5^{-},6^{+}) = \frac{\kappa^2}{4}\frac{\langle 5\slashed{3}6] ^4}{(5+6)^2}[\frac{1}{(5+3)^2-m^2}+\frac{1}{(5+4)^2-m^2}]\nonumber\\
%M(1,2,3^{+},4^{+}) = \frac{\kappa}{4}\frac{ \langle 3 \slashed{2} 4 ]^{4}}{[(3+4)^2]^{2}}[\frac{1}{(1+3)^2-m^2}+\frac{1}{(1+4)^2-m^2}]
\end{eqnarray}

\subsection{Spin orbit}
To begin, we consider the scattering of the scalar with a massive spin 1/2 fermion. For tree level scattering, we will use the usual Feynman rules. As before, a minimal coupling to gravity gives us
\begin{eqnarray}
iM(p_1,p_2,m_a) = \frac{-i\kappa}{2}[(p_1+p_2)_{\nu}\gamma_{\mu}+(p_1+p_2)_{\mu}\gamma_{\nu}-\eta_{\mu \nu}(\frac{1}{2}(\slashed{p_1}+ \slashed{p_2}) -m_a)]
\end{eqnarray}
where $p_1$ is incoming and $p_2$ is ougoing momentum of the fermion with mass $m_a$.
 %In principle, there are methods for using BCFW recursion for all massive external legs \cite{}, but for 2-2 scattering at tree level, there is no great computational advantage.
%%%%%%%%%%%%%%%%%%%%%%%%%%%%%%%%%%%%%%%%%%%%%%%%%%%%%%%%%%%%%
%\begin{figure}
%\centering
%\includegraphics[width=4in]{3point.jpg}
%\end{figure}
%%%%%%%%%%%%%%%%%%%%%%%%%%%%%%%%%%%%%%%%%%%%%%%%%%%%%%%%%%%%
On the other hand for loop calculations, generalized unitarity methods become invaluable and to use them we need the on shell three point amplitude.
\begin{equation}
M(1,2,5^{+})= \frac{\kappa}{2}\overline{u}(2)\gamma_{\mu}u(1)\frac{\langle r\gamma^{\mu}5]\langle r\slashed{1}5\rbrack}{\langle r5 \rangle^{2}}
\end{equation}
The expression for the graviton with negative helicity is similar with angles interchanged with square brackets. 
Using this seed we can use BCFW to construct the four point amplitudes.\\
\begin{eqnarray}
M(1,2,5^{-},6^{+}) = \frac{\kappa^2}{4}\overline{u}(2)\gamma_{\mu}u(1)\frac{\langle 5\gamma^{\mu}6]\langle 5\slashed{1}6\rbrack^{3}}{[(5+6)^2]^{2}}[\frac{1}{(1+5)^2-m_a^2}+\frac{1}{(1+6)^2-m_a^2}]
%M(1,2,3^{+},4^{+}) = \frac{\kappa}{4}m^{2}( \overline{u}(2)4\rbrack \langle 4u(1)\frac{[43]^{2}}{\langle 34 \rangle^{2}} + \overline{u}(2)\gamma_{\mu}u(1)\frac{\langle 4\gamma^{\mu}134\rbrack}{2\langle 34 \rangle^{3}} )\frac{1}{(1+3)^2-m^2}
\end{eqnarray}
\\
For calculating tree level scattering amplitudes, we use the graviton propagator in the harmonic or Feynman gauge. For all our calculations, we choose the incoming and outgoing scalar particles with rest mass $m_b$ to have momenta $p_3$ and $p_4$ respectively. The particles with non-zero spin with mass $m_a$ have momenta $p_1$ and $p_2$. In the center of mass frame 
\begin{eqnarray}
p_{1}^{\mu}=(E_{1},\vec{p}+\vec{q}/2), \ \ \ p_{2}^{\mu}=(E_{2},\vec{p}-\vec{q}/2), \ \ \ p_{3}^{\mu}=(E_{3},-\vec{p}-\vec{q}/2), \ \ \ p_{4}^{\mu}=(E_{4},-\vec{p}+\vec{q}/2)\nonumber
\end{eqnarray}
The non relativistic limit of this amplitude has been obtained in \cite{Holstein:2008sx}. In order to calculate the spin orbit piece upto 2.5 PN, we need to expand out the spin independent piece to 1PN order where we have kept only the classical contributions.

%\begin{eqnarray}
%M^{(1)}=-\frac{16\pi Gm_{a}m_{b}}{q^2}[-m_am_b\overline{u}(2)\gamma_{\mu}u(1) + \frac{s-m_a^2-m_b^2}{m_a}\overline{u}(2)\slashed{p_3}\gamma_{\mu}u(1)]
%\end{eqnarray}

%We now consider the non relativistic limit with a normalization factor $\frac{1}{4\sqrt{E_{1}E_{2}E_{3}E_{4}}}$  

\begin{eqnarray}
M&=&\frac{4\pi Gm_am_b}{\vec{q}^2}[\chi_{f}^{a\dagger}\chi_{i}^{a}(1+\frac{\vec{p}^{2}}{2m_{a}^{2}m_{b}^{2}}(3m_{a}^{2}+3m_{b}^{2}+8m_{a}m_{b}))\nonumber \\
&+&\frac{i\vec{S}\cdot (\vec{p} \times \vec{q})}{m_a^2m_b}[\frac{4m_a+3m_b}{2}+\frac{\vec{p}^2}{8m_a^2m_b}[8m_am_b-5m_b^2+18m_a^2]]
\end{eqnarray}  
where $\chi_{f}^{a}$ and $\chi_{i}^{a}$ are the spinors for the initial and final state of the fermion in the rest frame and $S^i = \chi_{f}^{a\dagger}\frac{\sigma^i}{2}\chi_{i}^{a}$
is the spin vector.

To extract out the effective potential we match this result onto an effective theory in which the graviton is integrated out.
\begin{eqnarray}
V_{si}(\vec{p},\vec{q})\psi^{\dagger}_{\vec{p}-\vec{q}/2}\psi_{\vec{p}+\vec{q}/2}\phi^{\dagger}_{-\vec{p}+\vec{q}/2}\phi_{-\vec{p}-\vec{q}/2}+V^j_{so}(\vec{p},\vec{q})S^j\phi^{\dagger}_{-\vec{p}+\vec{q}/2}\phi_{-\vec{p}-\vec{q}/2}
\end{eqnarray}
where $V_{si}$ is the spin independent and $V^j_{so}S^j$ is the spin orbit piece of the potential.
To get the complete spin orbit term at 2.5PN, we need to consider the scattering amplitude at one loop. 
Using generalized unitarity methods we can construct the one loop amplitude by sewing together the 4 point amplitudes for the scalar and fermion as shown in the Fig.1.

%%%%%%%%%%%%%%%%%%%%%%%%%%%%%%%%%%%%%%%%%%%%%%%%%%%%%%%%%%%%%
\begin{figure}
\centering
\includegraphics[width=4in]{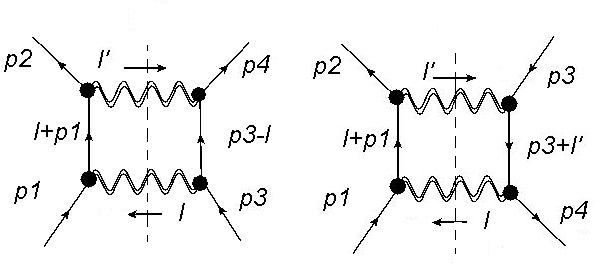}
\label{loop}
\caption{Fusing two tree level on-shell 4 point amplitudes}
\end{figure}
%%%%%%%%%%%%%%%%%%%%%%%%%%%%%%%%%%%%%%%%%%%%%%%%%%%%%%%%%%%%
\begin{eqnarray}
M(1,2,3,4)= \int{\frac{d^{4}l}{(2 \pi)^{4}}\frac{iM(1,2,l^{-},-l'^{+})iM(3,4,-l^{+},l'^{-}) + (+ \leftrightarrow -)}{l^{2}l'^{2}} }%\nonumber\\
%+\frac{iM(1,2,l^{-},-l'^{+})iM(3,4,-l^{+},l'^{-}) + (+ \leftrightarrow -)}{[(l+1)^{2}-m_{a}^{2}][(l'+3)^{2}-m_{b}^{2}]l^{2}l'^{2}}
\end{eqnarray}
%where $l'=l+q$, q being the momentum transfer betwen the two massive particles.\\

The basic idea is to simplify the numerator of the integrand by treating the gravitons(l,l') to be on shell in 4 D Space. After the simplification we will have a decomposition into standard scalar integrals. Using this we can accurately obtain the coefficients of those scalar integrals which contain $all$ the cut propagators. In this case we are going for a t-channel cut which involves a cut on the two massless graviton propagators. The only scalar integrals which give a classical contribution are those given by the triangle diagram with exactly one massive propagator. This means that t-channel cut is sufficient to calculate all the coefficients we need.\\
Moreover since we are using dimensional regularization, the loop integral in l is in d dimensions. But the reduction is much simpler in 4 dimensions and it is justified in this case since the errors produced are rational terms(polynomials) in the transfer momentum q which do not affect the long range classical result.
%\begin{eqnarray}
%M^{(2)}(1,2,3,4)= \frac{4G^{2}\pi^{2}m_{a}^{2}m_{b}^{2}}{q}[U_{a}(6m_{a}+6m_{b}) \nonumber\\
%+i\frac{E_{a}}{m_{a}^{2}m_{b}}(\frac{11(4m_{a}+3m_{b})}{4}+\frac{m_{a}m_{b}(4m_{a}+3m_{b})}{s-s_{0}})] 
%\end{eqnarray}
%We use the identity
%\begin{equation}
%\overline{u}(2)\gamma_{\mu}u(1)= \frac{1}{1-\frac{q^{2}}{4m_{a}^{2}}}[\frac{1^{\mu} + 2^{\mu}}{2m_{a}}\overline{u}(2)u(1)-\frac{i}{m_{a}^{2}}\epsilon_{\mu\beta\gamma\delta}q^{\beta}1^{\gamma}S_{a}^{\delta}]
%\end{equation}
%Where $S_{a}^{\mu}=\frac{1}{2}\overline{u}(2)\gamma_{5}\gamma^{\mu}u(1),\ \ \ U_{a}= \overline{u}(2)u(1)$ and 
%where we define
%\begin{eqnarray}
%U_{a}= \overline{u}(2)u(1)\\
%E_{a}=\epsilon_{\mu\beta\gamma\delta}1^{\mu}3^{\beta}q^{\gamma}S_{i}^{\delta}
%\end{eqnarray} 
%Here s is the usual mandelstam variable and $s_{0}$ is the threshold value $(m_{a}+m_{b})^{2}$. This is in complete agreement with the result obtained by holstein , ross \cite{}.\\
As before, we consider the non relativistic limit with a normalization factor $1/\sqrt{2E_12E_22E_32E_4}$ to give
%\begin{eqnarray}
%\overline{u}(2)u(1)\rightarrow \chi^{a \dagger}_{f}\chi^{a}_{i}-\frac{i}{2m_{a}^{2}}\vec{S_{a}}\cdot \vec{p}\times\vec{q}\\
%\epsilon_{\mu\beta\gamma\delta}1^{\mu}3^{\beta}q^{\gamma}S_{b}^{\delta} \rightarrow (m_{a}+m_{b})(1+\frac{\vec{p}^{2}}{2m_{a}m_{b}})\vec{S}_{b}\cdot \vec{p}\times\vec{q}\\
%S_{a}^{\mu} \rightarrow (\frac{\vec{S}_{a}\cdot \vec{p}}{m_{a}}
%,\nonumber\\
%\vec{S_{a}}[1+\frac{\vec{q}^{2}}{8m_{a}^{2}}]+\frac{1}{4m_{a}^{2}}[\vec{p}\vec{S}_{a}\cdot(2\vec{p})-\frac{\vec{q}}{2}\vec{S}_{a}\cdot\vec{q}-i\chi_{f}^{+}\chi_{i}(\vec{p}\times \vec{q})])
%\end{eqnarray}
\begin{eqnarray}
M^{(2)}&=&\frac{G^{2}\pi^{2}}{q}[m_am_b\chi_{f}^{a\dagger}\chi_{i}^{a}(6m_{a}+6m_{b})\nonumber\\
&+&i\vec{S}\cdot (\vec{p} \times \vec{q})[\frac{20m_a^3 +9m_b^3+53m_a^2m_b+41m_am_b^2}{2m_a(m_{a}+m_{b})} +\frac{m_a^2m_b^2(3m_b+4m_a)}{(m_a+m_b)p_0^2}]]\nonumber\\
\end{eqnarray} 
where $p_0^2 =\vec{p}^2 +\vec{q}^2/4$.
This result has a singular behavior in the limit $p_0\rightarrow 0$. To define a well behaved potential, we match onto our effective theory. This requires us to subtract out the iterated tree level potential from the loop scattering amplitude. To obtain the potential in position space, we Fourier transform the resulting coefficients for our effective therory with repect to the transfer momentum vector $\vec{q}$. We now match this non relativistic effective theory onto a point particle Hamiltonian by treating r as the conjugate position variable to the canonical momentum p. This choice of coordinate system is a specific one and hence makes the hamiltonian gauge dependent.  
	%This finally yields the results
	%\begin{eqnarray}
	%V_{si}(\vec{p},\vec{q})&=& \frac{Gm_am_b\pi}{\vec{q}^2}[(1+\frac{\vec{p}^{2}}{2m_{a}^{2}m_{b}^{2}}(3m_{a}^{2}+3m_{b}^{2}+8m_{a}m_{b}))]\nonumber\\
	%&-&\frac{G^{2}m_am_b\pi^{2}}{q}(m_{a}+m_{b})^2[1+\frac{m_am_b}{(m_a+m_b)^2}]\nonumber\\
	%V^j_{so}(\vec{p},\vec{q})S^j&=&\frac{iGS_{a}\cdot (p \times q)}{m_a^2m_b}[\frac{4m_a+3m_b}{2}+\frac{\vec{p}^2}{8m_a^2m_b}[8m_am_b-5m_b^2+18m_a^2]]\nonumber\\
	%&-&\frac{G^2\pi^2 \vec{S} \cdot (\vec{p} \times \vec{q})}{2qm_a(m_a+m_b)}[12 m_a^3+10m_b^3+45m_b^2m_a+41m_am_b^2]
	%\end{eqnarray}
%In position space, this leads to the Hamiltonian
\begin{eqnarray}
H&=& \frac{\vec{p}^2}{2m_a} + \frac{\vec{p}^2}{2m_b} - \frac{\vec{p}^4}{8m_a^3} - \frac{\vec{p}^4}{8m_b^3} -\frac{Gm_am_b}{r}[(1+\frac{\vec{p}^{2}}{2m_{a}^{2}m_{b}^{2}}(3m_{a}^{2}+3m_{b}^{2}+8m_{a}m_{b}))\nonumber\\
&+&\frac{\vec{S}\cdot (\vec{p} \times \vec{r})}{r^2m_a^2m_b}[\frac{4m_a+3m_b}{2}+\frac{\vec{p}^2}{8m_a^2m_b}[8m_am_b-5m_b^2+18m_a^2]]]\nonumber\\
	&+&\frac{G^2 \vec{S} \cdot (\vec{p} \times \vec{r})}{2r^4m_a(m_a+m_b)}[12 m_a^3+10m_b^3+45m_b^2m_a+41m_am_b^2]\nonumber\\
	&+&\frac{G^{2}}{2r^2}m_am_b(m_{a}+m_{b})[1+\frac{m_am_b}{(m_a+m_b)^2}]\nonumber
\end{eqnarray}
We are working in a frame in which the momentum($\vec{p}$) is directed transverse to $\vec{r}$ and we do not have a $\vec{p} \cdot \vec{r}$ term. In order to compare our result with existing literature, we choose a different coordinate system to express our result. This amounts to a canonical transformation of the Hamiltonian. The most general form of the generator to implement this transformation is 
\begin{eqnarray}
g=  a_1\frac{G(m_a+m_b)(\vec{p} \cdot \vec{r})}{r}+a_2\frac{G\vec{S} \cdot(\vec{p} \times \vec{r})(\vec{p} \cdot \vec{r})}{r^3}
\end{eqnarray}
This generates a correction to the Hamiltonian $\left\{g,H\right\}$. The choice of constants $a_1=	\frac{m_{a}m_{b}}{2(m_{a}+m_{b})^{2}}$ and $ a_2 =\frac{2m_a+m_b}{4m_a(m_a+m_b)}$ gives the result\\
\begin{eqnarray}
H_{int}&=& -\frac{Gm_am_b}{r}[(1+\frac{\vec{p}^{2}}{2m_{a}^{2}m_{b}^{2}}(3m_{a}^{2}+3m_{b}^{2}+7m_{a}m_{b})+\frac{(\vec{p} \cdot \vec{r})^2}{2m_am_b})]+\frac{G^{2}}{2r^2}m_am_b(m_{a}+m_{b})\nonumber\\
&-&\frac{G\vec{S}\cdot (\vec{p} \times \vec{r})}{r^3m_a}[\frac{4m_a+3m_b}{2}+\frac{\vec{p}^2}{8m_a^2m_b}[6m_am_b-5m_b^2+14m_a^2] +\frac{(\vec{p} \cdot \vec{r})^2}{4r^2m_a^2m_b}(6m_a+3m_b)]\nonumber\\
	&+&\frac{G^2 \vec{S} \cdot (\vec{p} \times \vec{r})}{2r^4m_a(m_a+m_b)}[12 m_a^3+10m_b^3+38m_bm_a^2+36m_am_b^2]
\end{eqnarray} 
\\
In this case, the spin independent result agrees with the EIH potential. The spin dependent piece agrees with the result obtained by Damour et al. \cite{Damour:2007nc} in the center of mass frame.

\subsection{Spin quadrupole}

The $S^2$ piece of the amplitude can be obtained by scattering the scalar with a massive spin 1 particle.
We begin with the Proca action for a massive particle of spin 1.
\begin{eqnarray}
S= \int{d^{4}x [-\frac{1}{4} G_{\mu \nu} G^{\mu \nu}+\frac{1}{2}m^2\phi^{\mu}\phi_{\mu}]}
\end{eqnarray}  
We consider a minimal coupling to gravity to determine the interaction. In this paper, we are interested only in the leading order $S^2$ piece, hence a tree level scattering amplitude is sufficient. This means that the massive particles are always on-shell and we can use equations of motion to simplify the stress energy tensor.
\begin{eqnarray}
T^{\mu \nu} &=& \frac{\kappa}{2} [\partial^{\mu} \phi_{\alpha} \partial^{\nu} \phi^{\alpha} +\partial_{\alpha}\phi^{\mu} \partial^{\alpha} \phi^{\nu} -\partial^{\mu} \phi^{\alpha} \partial_{\alpha} \phi^{\nu} - \partial^{\nu} \phi^{\alpha} \partial_{\alpha} \phi^{\mu} -m^2\phi^{\mu}\phi^{\nu}]\nonumber\\ 
&+&\frac{\kappa}{4} \eta^{\mu \nu}[-\partial_{\alpha}\phi_{\beta}\partial^{\alpha}\phi^{\beta} + \partial_{\alpha} \phi_{\beta}\partial^{\beta}\phi^{\alpha} +m^2\phi_{\alpha}\phi^{\alpha}]
\end{eqnarray}

Scattering this off a scalar with mass $m_b$ gives us the following scattering amplitude in the center of mass frame. 
\begin{eqnarray}
iM&=&-\frac{4\pi G m_am_b}{\vec{q^2}}[\epsilon^*(p_2) \cdot \epsilon(p_1) + \hat{p_1} \cdot \epsilon^*(p_2) \hat{p_2} \cdot \epsilon(p_1) \nonumber\\
&+&2[\hat{p_{2}} \cdot \epsilon(p_1) \hat{p_{3}} \cdot \epsilon^*(p_2) + \hat{p_{1}} \cdot \epsilon^*(p_2) \hat{p_{3}} \cdot \epsilon(p_1)-\hat{p_{2}} \cdot \epsilon(p_1) \hat{p_{1}} \cdot \epsilon^*(p_2)]]
\end{eqnarray}
where $\hat{p_1}=p_1/m_a,\ \hat{p_2}=p_2/m_a, \ \hat{p_3}=p_3/m_b, \ \hat{p_4}=p_4/m_b$.\\ 
We now consider the non-relativistic limit of this amplitude using the following approximations. 
\begin{eqnarray}
\epsilon^*(p_2) \cdot \epsilon(p_1) &\approx& -\hat{\epsilon_1} \cdot \hat{\epsilon^*_2}-\frac{1}{2m_a^2} \vec{q} \cdot \hat{\epsilon_1} \vec{q} \cdot \hat{\epsilon^*_2} - \frac{1}{2m_a^2}(q^ip^j-p^iq^j)\hat{\epsilon_1}^i \hat{\epsilon^*_2}^j\nonumber\\
\hat{p_{2}} \cdot \epsilon(p_1) \hat{p_{3}} \cdot \epsilon^*(p_2) + \hat{p_{1}} \cdot \epsilon^*(p_2) \hat{p_{3}} \cdot \epsilon(p_1) &\approx& -\frac{1}{m_a^2} \vec{q} \cdot \hat{\epsilon_1} \vec{q} \cdot \hat{\epsilon^*_2} - (\frac{1}{m_a^2}+\frac{1}{m_am_b})(q^ip^j-p^iq^j)\hat{\epsilon_1}^i \hat{\epsilon^*_2}^j\nonumber\\
\hat{p_{2}} \cdot \epsilon(p_1) \hat{p_{1}} \cdot \epsilon^*(p_2)&\approx& -\frac{1}{m_a^2} \vec{q} \cdot \hat{\epsilon_1} \vec{q} \cdot \hat{\epsilon^*_2}
\end{eqnarray}
where $\hat{\epsilon_i}$ is the polarization tensor of the spin 1 particle with momentum $p_i$ in the rest frame. 
This reduces the amplitude to the following compact form 
\\
\begin{eqnarray}
M \simeq \frac{4\pi G m_am_b}{\vec{q}^2}[\hat{\epsilon_1}^{i}\hat{\epsilon^*_2}^{i}-\frac{1}{m_a^2}q^{i}q^{j}\hat{\epsilon_1}^{i}\hat{\epsilon^*_2}^{j}+(\frac{3m_b+4m_a}{m_a^2m_b})q^ip^j(\hat{\epsilon_1}^{i}\hat{\epsilon^*_2}^{j}-\hat{\epsilon^*_2}^{i}\hat{\epsilon_1}^{j})]
\end{eqnarray}
\\
The effective potential between the two objects in position space is
%\begin{eqnarray}
%V(\vec{p},\vec{r})= \int{d^3q e^{-i\vec{q} \cdot \vec{r}}M(\vec{p},\vec{q})}
%\end{eqnarray}
%This gives us the potential
\begin{eqnarray}
V(\vec{p},\vec{r})= G m_am_b[-\frac{1}{r}\hat{\epsilon_1}^{ik}\hat{\epsilon^*_2}^{ki}-\frac{1}{m_a^2}(\frac{3r^ir^j}{r^5}-\frac{\delta^{ij}}{r^3})\hat{\epsilon_1}^{ik}\hat{\epsilon^*_2}^{kj}+i(\frac{3m_b+4m_a}{m_a^2m_b})\frac{r^i}{r^3}p^j(\hat{\epsilon_1}^{ik}\hat{\epsilon_2}^{kj}-\hat{\epsilon^*_2}^{ik}\hat{\epsilon_1}^{kj})]\nonumber
\end{eqnarray}\\
%which can be rewritten as \\
%\begin{eqnarray}
%V(\vec{p},\vec{r})= G m_1m_2[-\frac{1}{r}\hat{\epsilon_1}^{ik}\hat{\epsilon_2}^{ki}-\frac{3}{m_1^2}(3\hat{\epsilon_1}^{ik}\hat{\epsilon_2}^{kj}-\delta^{ij}\hat{\epsilon_1}^{ik}\hat{\epsilon_2}^{ki})\frac{r^ir^j}{r^5}+i(\frac{3m_2+4m_1}{m_1^2m_2})\frac{r^i}{r^3}p^j(\hat{\epsilon_1}^{ik}\hat{\epsilon_2}^{kj}-\hat{\epsilon_2}^{ik}\hat{\epsilon_1}^{kj})\nonumber
%\end{eqnarray}
%\\
In order to match this amplitude onto the effective theory, we need to consider the relevant operators that will appear in our EFT Lagrangian. In the rest frame of the particles, the only non trivial vector operator that is available is spin. This implies that any tensor constructed using the  polarization vectors has to map onto some linear combination of coreesponding tensors constructed using the spin vector and other invariant tensors. 
We can define the spin operators using the following identities,\\
\begin{eqnarray}
\hat{\epsilon_1}^{i}\hat{\epsilon^*_2}^{j}-\hat{\epsilon^*_2}^{i}\hat{\epsilon_1}^{j}= \frac{i}{2}\epsilon^{ijm}<s=1,m_2|S^m|s=1,m_1> \nonumber\\
\frac{3}{2}(\hat{\epsilon_1}^{i}\hat{\epsilon^*_2}^{j}+\hat{\epsilon^*_2}^{i}\hat{\epsilon_1}^{j})-\delta^{ij}\hat{\epsilon_1}^{k}\hat{\epsilon^*_2}^{k}=-<s=1,m_2|\frac{3}{2}(S^iS^j+S^jS^i)-\vec{S}^2\delta^{ij}|s=1,m_1>
\end{eqnarray}
\\
Here $m_1$ and $m_2$ are the z components of the spin in the initial and final state for the massive spin 1 particle.
%The final step involves matching onto an effective theory with an interaction Lagrangian of the form $$
%\begin{eqnarray}
%V(\vec{p},\vec{r})= G m_1m_2[-\frac{1}{r}-\frac{1}{2m_1^2r^3}(\frac{3(\vec{S} \cdot \vec{r})^2}{r^2}-\vec{S^2})+i(\frac{3m_2+4m_1}{2m_1^2m_2r^3})\vec{S} \cdot (\vec{r} \times \vec{p})]
%\end{eqnarray}
Apart form the minimal coupling, we can add other gauge invariant operators to the Proca Lagrangian. It turns out that the only relevant operator that we can add which has a nontrivial effects on the classical result is  
\begin{eqnarray}
L_{int}=\frac{C_1}{8}R_{\mu \nu \alpha \beta}G^{\mu \nu} G^{\alpha \beta}
\end{eqnarray}
This additional piece leaves the newtonian and spin orbit term unchanged, but alters the spin quadrupole term giving us the final result\\
\begin{eqnarray}
V(\vec{p},\vec{r})= G m_am_b[-\frac{1}{r}+(C_1-\frac{1}{2m_a^2})\frac{1}{r^3}(\frac{3(\vec{S} \cdot \vec{r})^2}{r^2}-\vec{S^2})+(\frac{3m_b+4m_a}{2m_a^2m_br^3})\vec{S} \cdot (\vec{r} \times \vec{p})] 
\end{eqnarray}
\\
Comparing with existing literature\cite{Hergt:2007ha} we see that this is the result for a black hole when $C_1=0$. This indicates that minimal coupling to gravity corresponds to a black hole structure.  This also demonstrates the universal form of the spin orbit term for the interaction between any two classical objects. 
The arbitrary coefficient $C_{1}$ allows us to account for any other massive classical object(e.g.  a neutron star). In order to determine this coefficient, we can do a matching procedure using any other spin dependent observable related to the star. For example, \cite{Chakrabarti:2013xza} uses an effective theory to model any star as a point source with finite size effects encoded into effective operators. This is essentially an expansion in multipolar degrees of freedom. The dynamics of these multipoles can be obtained by matching the gravitational field of the actual star with that of the effective point source. 
 %As for the spin 1/2 case, the operators $R\phi^{\mu}\phi_{\nu}$ and $R_{\mu \nu}\phi^{\mu} \phi^{\nu}$ do not affect the classical result.

\subsection{Spin octupole}
To derive the spin octupole Hamiltonian at leading order, we need to consider the scattering of a spin 2 particle. 
We begin with the Fierz Pauli action for a massive elementary particle with spin 2 \cite{Fierz:1939ix}
\begin{eqnarray}
S = \int{d^{4}x [-\frac{1}{2} \partial_{\lambda} \phi_{\mu \nu} \partial^{\lambda} \phi^{\mu \nu} + \partial_{\mu} \phi_{\nu \lambda} \partial^{\nu} \phi^{\mu \lambda} - \partial_{\mu}\phi^{\mu \nu}\partial_{\nu} \phi +\frac{1}{2}\partial_{\lambda}\phi\partial^{\lambda} \phi -\frac{1}{2}m^{2}(\phi_{\mu \nu} \phi^{\mu \nu} -\phi^2)]}\nonumber
\end{eqnarray}
here $\phi = \phi^{\mu}_{\mu}$ is the trace over the spin 2 tensor.\\
The equations of motion from this free field Lagrangian give a symmetric traceless rank 2 tensor which restricts the number of on shell modes to 5.
\begin{eqnarray}
\partial_{\mu} \phi^{\mu \nu} = 0 \nonumber\\
\phi = 0 \nonumber\\
(\partial^2 + m^2) \phi^{\mu \nu}=0
\end{eqnarray}
We consider a minimal coupling to gravity 
\begin{eqnarray}
S= \int{d^{4}x \sqrt{|g|}[-\frac{1}{2} \nabla_{\lambda} \phi_{\mu \nu} \nabla^{\lambda} \phi^{\mu \nu} + \nabla_{\mu} \phi_{\nu \lambda} \nabla^{\nu} \phi^{\mu \lambda} - \nabla_{\mu}\phi^{\mu \nu}\nabla_{\nu} \phi +\frac{1}{2}\nabla_{\lambda}\phi\nabla^{\lambda} \phi -\frac{1}{2}m^{2}(\phi_{\mu \nu} \phi^{\mu \nu} -\phi^2)]}\nonumber
\end{eqnarray} 
This gives us a symmetric and conserved stress energy tensor which we again simplify using the equations of motion. 
\begin{eqnarray}
T^{\gamma \delta} &=& -\partial^{\gamma}\phi_{\nu \lambda}\partial^{\nu}\phi^{\delta \lambda} - \partial^{\delta}\phi_{\nu \lambda}\partial^{\nu}\phi^{\gamma \lambda} +\partial_{\mu}\phi^{\nu \delta}\partial_{\nu}\phi^{\mu \gamma} +\frac{1}{2}\partial^{\gamma}\phi_{\nu \lambda}\partial^{\delta}\phi^{\lambda \nu} + \partial_{\mu}\phi^{\nu \gamma}\partial^{\mu}\phi^{\delta}_{\nu} - \partial_{\mu}\partial_{\nu}\phi^{\gamma \delta} \phi^{\mu \nu} \nonumber\\
&-& m^2\phi_{\mu}^{\gamma}\phi^{\mu \delta} +\frac{1}{2}\eta^{\gamma \delta}[-\frac{1}{2} \partial_{\lambda} \phi_{\mu \nu} \partial^{\lambda}\phi^{\mu \nu} + \partial_{\mu}\phi_{\nu \lambda}\partial^{\nu}\phi^{\mu \lambda} +\frac{1}{2}m^2\phi_{\mu \nu}\phi^{\mu \nu}]
\end{eqnarray}\\
%where we have used on-shell conditions since at tree level the spin-2 fields are always on shell.
%%%%%%%%%%%%%%%%%%%%%%%%%%%%%%%%%%%
%\begin{figure}
%\centering
%\includegraphics[width=1.5in]{spintwo.png}
%\caption{three point amplitude}
%\label{resum}
%\end{figure}
%%%%%%%%%%%%%%%%%%%%%%%%%%%%%%%%%%%%  
%This gives us the three point amplitude 
%\begin{eqnarray}
%iM(p_1,p_2,q)&=&[-p_{1}^{\gamma} p_{2}^{\alpha} \eta^{\beta \nu}\eta^{\mu \delta} -p_{2}^{\gamma}p_1^{\mu}\eta^{\beta \nu}\eta^{\alpha \delta} + p_1^{\mu}p_2^{\alpha}\eta^{\beta \delta}\eta^{\gamma \nu}+ \frac{1}{2} p_1^{\gamma}p_2^{\delta}\eta^{\mu \alpha}\eta^{\beta}+p_1 \cdot p_2 \eta^{\mu \alpha} \eta^{\nu \gamma}\eta^{\beta \delta}\nonumber\\
%&-&m^2\eta^{\mu \alpha} \eta^{\nu \gamma}\eta^{\beta \delta} +( \gamma \leftrightarrow \delta) ] + p_1^{\mu} p_2^{\nu} \eta^{\alpha \gamma} \eta^{\beta \delta} + p_2^{\alpha} p_2^{\beta} \eta^{\mu \gamma}\eta^{\nu \delta}\nonumber\\
%&+&\frac{1}{2} \eta^{\gamma \delta}[-p_1 \cdot p_2 \eta^{\mu \alpha} \eta^{\nu \beta}+2 p_1^{\mu}p_2^{\alpha}\eta^{\beta \nu}+ m^2\eta^{\mu \alpha} \eta^{\nu \beta}]
%\end{eqnarray}
%%%%%%%%%%%%%%%%%%%%%%%%%%%%%%%%%%%
%\begin{figure}
%\centering
%\includegraphics[width=2.5in]{scatter.png}
%\caption{scatter}
%\label{scatter}
%\end{figure}
%%%%%%%%%%%%%%%%%%%%%%%%%%%%%%%%%%%%  
We now consider leading order elastic scattering amplitude between a massive spin 2 particle and a massive scalar. \\
\begin{eqnarray}
M= \frac{4\pi G m_am_b}{\vec{q}^2}[\epsilon(p_1)^{\mu \nu}\epsilon^*(p_2)_{\mu \nu} -4\epsilon(p_1)^{\alpha \beta}\epsilon^*(p_2)_{\beta \nu}(\hat{p_2}_{\alpha}\hat{p_3}^{\nu}+\hat{p_3}_{\alpha}\hat{p_1}^{\nu})\nonumber\\
+2\epsilon(p_1)^{\alpha \beta}\epsilon^*(p_2)^{\mu \nu}(2\hat{p_2}_{\alpha}\hat{p_3}_{\beta}\hat{p_1}_{\mu}\hat{p_3}_{\nu}+\hat{p_3}_{\alpha}\hat{p_3}_{\beta}\hat{p_1}_{\mu}\hat{p_1}_{\nu}+\hat{p_2}_{\alpha}\hat{p_2}_{\beta}\hat{p_3}_{\mu}\hat{p_3}_{\nu})] 
\end{eqnarray}
In order to extract out the effective potential, we take the non relativistic limit of this amplitude. This can be done using the following approximations
\begin{eqnarray}
\epsilon(p_1)^{\mu \nu}\epsilon^*(p_2)_{\mu \nu} \simeq \hat{\epsilon_1}^{ik}\hat{\epsilon^*_2}^{ki}+[\frac{1}{m_a^2}q^{i}q^{j}-\frac{1}{m_1^2}(q^ip^j-p^iq^j)]\hat{\epsilon_1}^{ik}\hat{\epsilon^*_2}^{kj}\nonumber\\
+[\frac{1}{2m_a^4}q^iq^j(p^kq^l-p^lq^k)+\frac{1}{4m_a^4}q^{i}q^{j}q^{k}q^{l}]\hat{\epsilon_1}^{ik}\hat{\epsilon^*_2}^{jl}
\end{eqnarray}
\\
where repeated indices are summed over and are all spatial.  $\hat{\epsilon_1}^{ij}, \hat{\epsilon^*_2}^{kl}$ are the polarization tensors in the rest frame.
\begin{eqnarray}
\epsilon(p_1)^{\alpha \beta}\epsilon^*(p_2)_{\beta \nu}(\hat{p_2}_{\alpha}\hat{p_3}^{\nu}+\hat{p_3}_{\alpha}\hat{p_1}^{\nu}) \simeq  [\frac{1}{m_a^2}q^{i}q^{j} -(\frac{1}{m_a^2}+\frac{1}{m_am_b})(q^ip^j-p^iq^j)]\hat{\epsilon_1}^{ik}\hat{\epsilon^*_2}^{kj}\nonumber\\
+[(\frac{1}{m_a^4}+\frac{1}{2m_a^3m_b})q^iq^j(p^kq^l-p^lq^k)+\frac{1}{2m_a^4}q^{i}q^{j}q^{k}q^{l}]\hat{\epsilon_1}^{ik}\hat{\epsilon^*_2}^{jl}
\end{eqnarray}

\begin{eqnarray}
\epsilon(p_1)^{\alpha \beta}\epsilon^*(p_2)^{\mu \nu}(2\hat{p_2}_{\alpha}\hat{p_3}_{\beta}\hat{p_1}_{\mu}\hat{p_3}_{\nu}+\hat{p_3}_{\alpha}\hat{p_3}_{\beta}\hat{p_1}_{\mu}\hat{p_1}_{\nu}+\hat{p_2}_{\alpha}\hat{p_2}_{\beta}\hat{p_3}_{\mu}\hat{p_3}_{\nu}) \nonumber\\
\simeq  [2(\frac{1}{m_a^4}+\frac{1}{m_am_b})q^iq^j(p^kq^l-p^lq^k)+\frac{1}{m_a^4}q^{i}q^{j}q^{k}q^{l}]\hat{\epsilon_1}^{ik}\hat{\epsilon^*_2}^{jl}
\end{eqnarray}
%\begin{eqnarray}
%\epsilon(p_1)^{\alpha \beta}\epsilon(p_2)_{\beta \nu}{p_2}_{\alpha}\hat{p_1}^{\nu} \simeq \frac{1}{m_1^2}q^{i}q^{j}\hat{\epsilon_1}^{ik}\hat{\epsilon_2}^{kj}+\frac{1}{2m_1^4}q^iq^j(p^kq^l-p^lq^k)\hat{\epsilon_1}^{ik}\hat{\epsilon_2}^{jl}+\frac{1}{2m_1^4}q^{i}q^{j}q^{k}q^{l}\hat{\epsilon_1}^{ik}\hat{\epsilon_2}^{jl}
%\end{eqnarray} 

This reduces the amplitude to the following compact form
\\
\begin{eqnarray}
M \simeq \frac{4\pi G m_am_b}{\vec{q}^2}[\hat{\epsilon_1}^{ik}\hat{\epsilon^*_2}^{ki}-\frac{3}{m_a^2}q^{i}q^{j}\hat{\epsilon_1}^{ik}\hat{\epsilon^*_2}^{kj}+(\frac{3m_b+4m_a}{m_a^2m_b})q^ip^j(\hat{\epsilon_1}^{ik}\hat{\epsilon^*_2}^{kj}-\hat{\epsilon_2}^{ik}\hat{\epsilon_1}^{kj})\nonumber\\
+(\frac{1}{2m_a^4}+\frac{2}{m_a^3m_b})q^iq^jp^kq^l(\hat{\epsilon_1}^{ik}\hat{\epsilon^*_2}^{jl}-\hat{\epsilon^*_2}^{ik}\hat{\epsilon_1}^{jl})+\frac{1}{4m_a^4}q^{i}q^{j}q^{k}q^{l}\hat{\epsilon_1}^{ik}\hat{\epsilon^*_2}^{jl}]
\end{eqnarray}
\\
%As before, the effective potential between the two objects is now obtained by simply considering a fourier transform with respect to the transfer momentum $\vec{q}$.
which in turn gives us the potential
\begin{eqnarray}
V(\vec{p},\vec{r})&=& G m_am_b[-\frac{1}{r}\hat{\epsilon_1}^{ik}\hat{\epsilon^*_2}^{ki}-\frac{3}{m_a^2}(\frac{3r^ir^j}{r^5}-\frac{\delta^{ij}}{r^3})\hat{\epsilon_1}^{ik}\hat{\epsilon^*_2}^{kj}+i(\frac{3m_b+4m_a}{m_a^2m_b})\frac{r^i}{r^3}p^j(\hat{\epsilon_1}^{ik}\hat{\epsilon^*_2}^{kj}-\hat{\epsilon^*_2}^{ik}\hat{\epsilon_1}^{kj})\nonumber\\
&+&3i(\frac{1}{2m_a^4}+\frac{2}{m_a^3m_b})p^i(\frac{\delta^{kl}r^j}{r^5}+\frac{\delta^{jl}r^k}{r^5}+\frac{\delta^{kj}r^l}{r^5}-5\frac{r^kr^jr^l}{r^7})(\hat{\epsilon_1}^{ik}\hat{\epsilon^*_2}^{jl}-\hat{\epsilon_2}^{il}\hat{\epsilon_1}^{jk})\nonumber\\
&-&\hat{\epsilon_1}^{ik}\hat{\epsilon^*_2}^{jl}\frac{3}{4m_a^4}(\frac{\delta^{ij}\delta^{kl}}{r^5}+\frac{\delta^{ik}\delta^{jl}}{r^5}+\frac{\delta^{il}\delta^{jk}}{r^5}\nonumber\\
&-&\frac{5}{r^7}(r^ir^j\delta^{kl}+r^ir^k\delta^{jl}+r^jr^l\delta^{ik}+r^jr^k\delta^{il}+r^jr^l\delta^{kj}+r^kr^l\delta^{ij})+35\frac{r^ir^jr^kr^l}{r^9})]
\end{eqnarray}

%This can be rewritten as 
%\begin{eqnarray}
%V(\vec{p},\vec{r})&=& G m_1m_2[-\frac{1}{r}\hat{\epsilon_1}^{ik}\hat{\epsilon_2}^{ki}-\frac{3}{m_1^2}(3\hat{\epsilon_1}^{ik}\hat{\epsilon_2}^{kj}-\delta^{ij}\hat{\epsilon_1}^{ik}\hat{\epsilon_2}^{ki})\frac{r^ir^j}{r^5}+i(\frac{3m_2+4m_1}{m_1^2m_2})\frac{r^i}{r^3}p^j(\hat{\epsilon_1}^{ik}\hat{\epsilon_2}^{kj}-\hat{\epsilon_2}^{ik}\hat{\epsilon_1}^{kj})\nonumber\\
%&+&3(\frac{1}{2m_1^4}+\frac{2}{m_1^3m_2})\frac{p^kr^ir^jr^l}{r^7}(\delta^{il}(\hat{\epsilon_1}^{hk}\hat{\epsilon_2}^{jh}-\hat{\epsilon_2}^{hk}\hat{\epsilon_1}^{jh})+\delta^{jl}%(\hat{\epsilon_1}^{hh}\hat{\epsilon_2}^{ik}-\hat{\epsilon_2}^{hh}\hat{\epsilon_1}^{ik})\nonumber\\
%&+&\delta^{ij}(\hat{\epsilon_1}^{hk}\hat{\epsilon_2}^{lh}-\hat{\epsilon_2}^{hk}\hat{\epsilon_1}^{hl})-5(\hat{\epsilon_1}^{ik}\hat{\epsilon_2}^{jl}-\hat{\epsilon_2}^{ik}\hat{\epsilon_1}^{jl}))\nonumber\\
%&+&\frac{r^ir^jr^kr^l}{r^9}\frac{3}{4m_1^4}(\hat{\epsilon_1}^{hm}\hat{\epsilon_2}^{hm}\delta^{ij}\delta^{kl}+\hat{\epsilon_1}^{hh}\hat{\epsilon_2}^{mm}\delta^{ik}\delta^{jl}+\hat{\epsilon_1}^{hm}\hat{\epsilon_2}^{hm}\delta^{il}\delta^{jk}\nonumber\\
%&-&5(\hat{\epsilon_1}^{ih}\hat{\epsilon_2}^{jh}\delta^{kl}+\hat{\epsilon_1}^{ik}\hat{\epsilon_2}^{hh}\delta^{jl}+\hat{\epsilon_1}^{hh}\hat{\epsilon_2}^{jl}\delta^{ik}+\hat{\epsilon_1}^{hk}\hat{\epsilon_2}^{hl}\delta^{il}+\hat{\epsilon_1}^{ih}\hat{\epsilon_2}^{jh}\delta^{kj}+\hat{\epsilon_1}^{hk}\hat{\epsilon_2}^{hl}\delta^{ij})+35\hat{\epsilon_1}^{ik}\hat{\epsilon_2}^{jl})]\nonumber\\
%\end{eqnarray}
As for the case of spin 1, now match onto the spin operators. The easiest way to do this is for the case of spin 2 is to match the coefficients of irreducible tensor structures.
\begin{eqnarray}
\hat{\epsilon_1}^{ik}\hat{\epsilon^*_2}^{kj}-\hat{\epsilon^*_2}^{ik}\hat{\epsilon_1}^{kj}&=& \frac{-i}{2}\epsilon^{ijm}<s=2,m_2|S^m|s=2,m_1>\nonumber\\
\frac{3}{2}(\hat{\epsilon_1}^{ik}\hat{\epsilon^*_2}^{kj}+\hat{\epsilon^*_2}^{ik}\hat{\epsilon_1}^{kj})-\delta^{ij}\hat{\epsilon_1}^{ik}\hat{\epsilon_2}^{ki}&=&-\frac{1}{6}<s=2,m_2|\frac{3}{2}(S^iS^j+S^jS^i)-\vec{S}^2\delta^{ij}|s=2,m_1>\nonumber
\end{eqnarray}
The identities for $S^3$ and $S^4$ operators is more involved due to the multitude of non equivalent structures possible.
\begin{eqnarray}
&&\Big\{2\delta^{jk}(\hat{\epsilon_1}^{hi}\hat{\epsilon^*_2}^{lh}-\hat{\epsilon^*_2}^{hi}\hat{\epsilon_1}^{lh})-5 (\hat{\epsilon_1}^{ij}\hat{\epsilon^*_2}^{kl}-\hat{\epsilon^*_2}^{ik}\hat{\epsilon_1}^{jl})\Big\} +(j \leftrightarrow l)+ (k \leftrightarrow l)\nonumber\\
%+\delta^{jl}(\hat{\epsilon_1}^{ih}\hat{\epsilon_2}^{hk}-\hat{\epsilon_2}^{-h}\hat{\epsilon_1}^{hk})+\delta^{kl}(\hat{\epsilon_1}^{hi}\hat{\epsilon_2}^{jh}-\hat{\epsilon_2}^{hi}\hat{\epsilon_1}^{hj})-5(\hat{\epsilon_1}^{ik}\hat{\epsilon_2}^{jl}-\hat{\epsilon_2}^{ik}\hat{\epsilon_1}^{jl})\nonumber\\
&=&\frac{1}{18}\langle s=2,m_2|i\Big\{\delta^{jk}[3\epsilon^{ilm}S^m\vec{S}^2-S^i\epsilon^{alm}S^mS^a-\epsilon^{alm}S^mS^aS^i]\nonumber\\
&-&\frac{5}{2}[\epsilon^{ilm}S^mS^jS^k+S^l\epsilon^{ijm}S^mS^k+S^lS^j\epsilon^{ikm}S^m+\epsilon^{ikm}S^mS^lS^j+S^k\epsilon^{ilm}S^mS^j+S^kS^l\epsilon^{ijm}S^m]\Big\}\nonumber\\
&+&(j \leftrightarrow l)+ (k \leftrightarrow l)|s=2,m_1 \rangle\nonumber
\end{eqnarray}
\\
\begin{eqnarray}
&&\Big\{\hat{\epsilon_1}^{hm}\hat{\epsilon^*_2}^{hm}(\delta^{ij}\delta^{kl}+\delta^{il}\delta^{jk})-5(\hat{\epsilon_1}^{ih}\hat{\epsilon^*_2}^{jh}\delta^{kl}+\hat{\epsilon_1}^{hk}\hat{\epsilon^*_2}^{hl}\delta^{il}+\hat{\epsilon_1}^{ih}\hat{\epsilon^*_2}^{jh}\delta^{kj}+\hat{\epsilon_1}^{hk}\hat{\epsilon^*_2}^{hl}\delta^{ij})+35\hat{\epsilon_1}^{ik}\hat{\epsilon^*_2}^{jl}\Big\}\nonumber\\
 &+&\text{all permutations of } {i,j,k,l}\nonumber\\
&=&\frac{1}{6}\langle s=2,m_2|\Big\{(\vec{S}^2\vec{S}^2\delta^{ij}\delta^{kl}+\vec{S^a}\vec{S^b}\vec{S^a}\vec{S^b}\delta^{ik}\delta^{jl}+\vec{S}^2\vec{S}^2\delta^{il}\delta^{jk})\nonumber\\
&-&5(\vec{S^2}S^iS^j\delta^{kl}+S^iS^aS^kS^a\delta^{jl}+S^aS^jS^aS^l\delta^{ik}+S^aS^jS^kS^a\delta^{il}+\vec{S^2}S^iS^l\delta^{kj}+\vec{S^2}S^kS^l\delta^{ij})\nonumber\\
&+&35S^iS^jS^kS^l \Big\}+\text{all permutations of } {i,j,k,l}|s=2,m_1\rangle \nonumber\\
\end{eqnarray}
%which finally gives us the result
%\begin{eqnarray}
%V(\vec{p},\vec{r})= G m_1m_2[-\frac{1}{r}-\frac{1}{2m_1^2r^3}(\frac{3(\vec{S} \cdot \vec{r})^2}{r^2}-\vec{S^2})+i(\frac{3m_2+4m_1}{2m_1^2m_2r^3})\vec{S} \cdot (\vec{r} \times \vec{p}) \nonumber\\
%+3(\frac{1}{2m_1^4}+\frac{2}{m_1^3m_2})\vec{S} \cdot (\vec{r} \times \vec{p})(\frac{(\vec{S} \cdot \vec{r})^2}{r^2} + \vec{S}^2)\nonumber\\
%+\frac{3}{24m_1^4r^5}(3(\vec{S^2})^2-30\frac{\vec{S^2}(\vec{S} \cdot \vec{r})^2}{r^7}+35\frac{(\vec{S} \cdot \vec{r})^2)}{r^9}]\nonumber\\
We can also add three relevant gauge invariant operators :\\
\begin{eqnarray}
L_{int}=\frac{C_1}{8m_a^2}R_{\mu \nu \alpha \beta}U^{\mu \nu \gamma}U^{\alpha \beta}_{\gamma} +  C_2R_{\alpha \beta \gamma \rho}(\phi^{\alpha \gamma}\phi^{\beta \rho} - \phi^{\beta \gamma}\phi^{\alpha \delta})+\frac{C_3}{2m_a^2}R_{\mu \nu \alpha \beta}\partial^{\mu}\phi^{\rho \alpha}\partial_{\rho}\phi^{\nu \beta}
%C_1R_{\mu \nu}\phi^{\mu \alpha} \phi^{\nu}_{\alpha}+C_2R_{\alpha \beta \gamma \rho}(\phi^{\alpha \gamma}\phi^{\beta \rho} - \phi^{\beta \gamma}\phi^{\alpha \delta})
\end{eqnarray}
where $U^{\mu \nu \gamma} =\partial^{\mu} \phi^{\nu \gamma} -\partial^{\nu} \phi^{\mu \gamma}$.
These additional pieces leave the newtonian and spin orbit term unchanged, but alter the spin quadrupole and octupole terms giving us the result\\
\begin{eqnarray}
V(\vec{p},\vec{r})= G m_am_b[-\frac{1}{r}+[(C_1+\frac{1}{2}+\frac{C_2}{3})\frac{1}{m_a^2r^3}](\frac{3(\vec{S} \cdot \vec{r})^2}{r^2}-\vec{S^2})+(\frac{3m_b+4m_a}{2m_a^2m_br^3})\vec{S} \cdot (\vec{r} \times \vec{p}) \nonumber\\
+\frac{1}{2r^4}\Big\{ (C_3+4C_1)(\frac{1}{m_a^2}+\frac{1}{m_am_b})+(\frac{1}{2m_a^4}+\frac{2}{m_a^3m_b})\Big\}\vec{S} \cdot (\vec{r} \times \vec{p})( \vec{S}^2-5\frac{(\vec{S} \cdot \vec{r})^2}{r^2})\nonumber\\
-[\frac{C_1}{m_a^4r^5}+\frac{4C_2+1}{8m_a^4r^5}](3(\vec{S^2})^2-30\frac{\vec{S^2}(\vec{S} \cdot \vec{r})^2}{r^2}+35\frac{(\vec{S} \cdot \vec{r})^2)}{r^4})]\nonumber\\
\end{eqnarray}
 This gives us the result for the missing $H_{S^4}$ Hamiltonian for a compact star.
\begin{eqnarray}
H_{S^4}= -[\frac{C_1}{m_a^4r^5}+\frac{4C_2+1}{8m_a^4r^5}](3(\vec{S^2})^2-30\frac{\vec{S^2}(\vec{S} \cdot \vec{r})^2}{r^2}+35\frac{(\vec{S} \cdot \vec{r})^2)}{r^4})]
\end{eqnarray}
\\

As before, we recover the universal form of the spin orbit piece. 
 Since we have three additional operators for spin 2 case, we also have arbitary coefficients for the $S^2$, $S^3$ and $S^4$ pieces. The limit for the black hole is obtained for $C_1=0, C_2=0, C_3=0$ \cite{Hergt:2008jn}, which again demonstrates that a minimal coupling to the graviton implies a black hole. The wilson coefficients for these operators can be obtained form a matching procedure with any other spin depenedent observable. The results for $S^3$ hamiltonian was derived in the limit of a black hole (\cite{Hergt:2007ha},\cite{Hergt:2008jn}). An attempt to derive the quartic spin Hamiltonian for a black hole was made in \cite{Hergt:2008jn} but was found to be inconsistent with the results of \cite{Steinhoff:2012rw} which computes the binding energy of a test particle in the extreme mass ration in a cicular orbit with the spin of the massive star aligned perpendicular to the orbit. In this limit, the Hamiltonian above reduces to
As a consistency check we compare this result with \cite{Steinhoff:2012rw} which calculates the binding energy  
\begin{eqnarray}
H_{S^4}= -\frac{3G m_b}{8m_a^3r^5}(\vec{S^2})^2
\end{eqnarray}
A comparison with the result \cite{Steinhoff:2012rw} gives a match for the binding energy. 

\section{Kerr metric}
As another consistency check we can easily obtain the Kerr metric to leading power in G and upto 4th order in spin using the calculation done so far. 
We consider the world line action of a probe particle in a Kerr background field.
\begin{eqnarray}
S = -m_b\int{dt\sqrt{g_{00}+g_{0i}v_i +g_{ij}v_{ij}} }
\end{eqnarray}
For the leading order spin dependent pieces, this gives us the result
\begin{eqnarray}
g_{00} = 1 +2G m_a[-\frac{1}{r}+\frac{1}{2m_a^2r^3}(\frac{3(\vec{S} \cdot \vec{r})^2}{r^2}-\vec{S^2})-\frac{1}{8m_a^4r^5}(3(\vec{S^2})^2-30\frac{\vec{S^2}(\vec{S} \cdot \vec{r})^2}{r^2}+35\frac{(\vec{S} \cdot \vec{r})^4)}{r^4})]\nonumber\\
%g_{0i}=G m_1[(\frac{2}{m_1r^3}) (\vec{r} \times\vec{S})^i +3(\frac{2}{m_1^3}) (\vec{r} \times \vec{S})^i(\frac{(\vec{S} \cdot \vec{r})^2}{r^2} + \vec{S}^2)]\nonumber\\
\end{eqnarray}
Comparing $g_{00}$ with the corresponding result for the Kerr metric in harmonic co-ordinates \cite{Aguirregabiria:2001vk} again confirms the $S^4$ Hamiltonian peice.

\section{Summary and outlook}
We have used modern methods of amplitude computation combined with effective field theory techniques to obtain spin dependent Hamiltonians for a binary inspiralling system in the post newtonian approximation. The use of on shell methods substantially reduces the effort of computing loop diagrams. We have also shown how the idea of treating gravity as spin 2 massless particle provides a natural way of obtaining higher order spin corrections for arbitary classical objects. The possible gauge invariant interaction operators that we can write down, automatically account for any spinning classical objects including a black hole. Using a massive spin 2 particle scattering at tree level, we were able to obtain the $S^3$ interaction for an arbitrary object in terms of coefficients which depend on the specific equation of state for a star, which was till now known only for a black hole. We were also able to calculate in a simple manner the hitherto unknown $S^4$ Hamiltonian and show that three independent operators are needed to account for other stellar equations of state upto $4^{th}$ order in the PN expansion. What is really interesting, is the universality of interaction terms that appear as we move to particles of higher spin. Also, a curious fact is revealed that the minimal coupling of a massive elementary particle to gravity automatically accounts for any spin dependent effects of a black hole. In principle, all the spin dependent Hamiltonians upto 4PN order can be obtained by considering loop corrections for scattering of two spin 1 particles. The spin 2 particle scattering is required only at tree level.\\

Note added: As this paper was being finalized, another paper appeared\cite{Levi:2014gsa} which also investigates the cubic and quartic spin Hamiltonians using the NRGR  formalism which uses the traditional Effective field theory approach.

%%%%%%%%%%%%%%%%%%%%%%%%%%%%%%%%%%%%%%%%%%%%%%%%%%%%%%%%%%%%%
%%%%%%%%%%%%%%%%%%%%%%%%%%%%%%%%%%%%%%%%%%%%%%%%%%%%%%%%%%%%
\section{Acknowledgements}
I thank Ira Rothstein for his help on several conceptual aspects of this project and for comments on this manuscript. This work is supported by DOE contracts DOE-ER-40682-143 and DEACO2-C6H03000.


\begin{thebibliography}{99}
%\cite{Fierz:1939ix}

%\cite{Aasi:2012rja}
\bibitem{Aasi:2012rja} 
  J.~Aasi {\it et al.}  [LIGO Scientific and Virgo Collaborations],
  %``Search for Gravitational Waves from Binary Black Hole Inspiral, Merger and Ringdown in LIGO-Virgo Data from 2009-2010,''
  Phys.\ Rev.\ D {\bf 87}, 022002 (2013)
  [Phys.\ Rev.\ D {\bf 87}, 022002 (2013)]
  [arXiv:1209.6533 [gr-qc]].
  %%CITATION = ARXIV:1209.6533;%%
  %40 citations counted in INSPIRE as of 22 Sep 2014

%\cite{Abadie:2010cf}
\bibitem{Abadie:2010cf} 
  J.~Abadie {\it et al.}  [LIGO Scientific and Virgo Collaborations],
  %``Predictions for the Rates of Compact Binary Coalescences Observable by Ground-based Gravitational-wave Detectors,''
  Class.\ Quant.\ Grav.\  {\bf 27}, 173001 (2010)
  [arXiv:1003.2480 [astro-ph.HE]].
  %%CITATION = ARXIV:1003.2480;%%
  %360 citations counted in INSPIRE as of 22 Sep 2014



\bibitem{Fierz:1939ix} 
  M.~Fierz and W.~Pauli,
  %``On relativistic wave equations for particles of arbitrary spin in an electromagnetic field,''
  Proc.\ Roy.\ Soc.\ Lond.\ A {\bf 173}, 211 (1939).
  %%CITATION = PRSLA,A173,211;%%
  %845 citations counted in INSPIRE as of 31 Aug 2014

%\cite{Hergt:2012zx}
\bibitem{Hergt:2012zx} 
S.~Hergt, J.~Steinhoff and G.~Schaefer,
  %``On the comparison of results regarding the post-Newtonian approximate treatment of the dynamics of extended spinning compact binaries,''
  J.\ Phys.\ Conf.\ Ser.\  {\bf 484}, 012018 (2014)
  [arXiv:1205.4530 [gr-qc]].
  %%CITATION = ARXIV:1205.4530;%%
  %3 citations counted in INSPIRE as of 21 Sep 2014

%\cite{Damour:2007nc}
\bibitem{Damour:2007nc} 
  T.~Damour, P.~Jaranowski and G.~Schaefer,
  %``Hamiltonian of two spinning compact bodies with next-to-leading order gravitational spin-orbit coupling,''
  Phys.\ Rev.\ D {\bf 77}, 064032 (2008)
  [arXiv:0711.1048 [gr-qc]].
  %%CITATION = ARXIV:0711.1048;%%
  %86 citations counted in INSPIRE as of 21 Sep 2014

%\cite{Steinhoff:2008zr}
\bibitem{Steinhoff:2008zr} 
  J.~Steinhoff, G.~Schaefer and S.~Hergt,
  %``ADM canonical formalism for gravitating spinning objects,''
  Phys.\ Rev.\ D {\bf 77}, 104018 (2008)
  [arXiv:0805.3136 [gr-qc]].
  %%CITATION = ARXIV:0805.3136;%%
  %49 citations counted in INSPIRE as of 21 Sep 2014

%\cite{Steinhoff:2007mb}
\bibitem{Steinhoff:2007mb} 
  J.~Steinhoff, S.~Hergt and G.~Schaefer,
  %``On the next-to-leading order gravitational spin(1)-spin(2) dynamics,''
  Phys.\ Rev.\ D {\bf 77}, 081501 (2008)
  [arXiv:0712.1716 [gr-qc]].
  %%CITATION = ARXIV:0712.1716;%%
  %66 citations counted in INSPIRE as of 21 Sep 2014
	
	%\cite{Hergt:2010pa}
\bibitem{Hergt:2010pa} 
  S.~Hergt, J.~Steinhoff and G.~Schaefer,
  %``Reduced Hamiltonian for next-to-leading order Spin-Squared Dynamics of General Compact Binaries,''
  Class.\ Quant.\ Grav.\  {\bf 27}, 135007 (2010)
  [arXiv:1002.2093 [gr-qc]].
  %%CITATION = ARXIV:1002.2093;%%
  %29 citations counted in INSPIRE as of 21 Sep 2014
	
	%\cite{Levi:2011eq}
\bibitem{Levi:2011eq} 
  M.~Levi,
  %``Binary dynamics from spin1-spin2 coupling at fourth post-Newtonian order,''
  Phys.\ Rev.\ D {\bf 85}, 064043 (2012)
  [arXiv:1107.4322 [gr-qc]].
  %%CITATION = ARXIV:1107.4322;%%
  %36 citations counted in INSPIRE as of 08 Nov 2014

%\cite{Goldberger:2004jt}
\bibitem{Goldberger:2004jt} 
  W.~D.~Goldberger and I.~Z.~Rothstein,
  %``An Effective field theory of gravity for extended objects,''
  Phys.\ Rev.\ D {\bf 73}, 104029 (2006)
  [hep-th/0409156].
  %%CITATION = HEP-TH/0409156;%%
  %150 citations counted in INSPIRE as of 22 Sep 2014

	%\cite{Porto:2010tr}
\bibitem{Porto:2010tr} 
  R.~A.~Porto,
  %``Next to leading order spin-orbit effects in the motion of inspiralling compact binaries,''
  Class.\ Quant.\ Grav.\  {\bf 27}, 205001 (2010)
  [arXiv:1005.5730 [gr-qc]].
  %%CITATION = ARXIV:1005.5730;%%
  %42 citations counted in INSPIRE as of 21 Sep 2014
	
	%\cite{Porto:2008jj}
\bibitem{Porto:2008jj} 
  R.~A.~Porto and I.~Z.~Rothstein,
  %``Next to Leading Order Spin(1)Spin(1) Effects in the Motion of Inspiralling Compact Binaries,''
  Phys.\ Rev.\ D {\bf 78}, 044013 (2008)
  [Erratum-ibid.\ D {\bf 81}, 029905 (2010)]
  [arXiv:0804.0260 [gr-qc]].
  %%CITATION = ARXIV:0804.0260;%%
  %78 citations counted in INSPIRE as of 21 Sep 2014
	
	%\cite{Porto:2008tb}
\bibitem{Porto:2008tb} 
  R.~A.~Porto and I.~Z.~Rothstein,
  %``Spin(1)Spin(2) Effects in the Motion of Inspiralling Compact Binaries at Third Order in the Post-Newtonian Expansion,''
  Phys.\ Rev.\ D {\bf 78}, 044012 (2008)
  [Erratum-ibid.\ D {\bf 81}, 029904 (2010)]
  [arXiv:0802.0720 [gr-qc]].
  %%CITATION = ARXIV:0802.0720;%%
  %83 citations counted in INSPIRE as of 21 Sep 2014
	
	%\cite{Rothstein:2003mp}
\bibitem{Rothstein:2003mp} 
  I.~Z.~Rothstein,
  %``TASI lectures on effective field theories,''
  hep-ph/0308266.
  %%CITATION = HEP-PH/0308266;%%
  %58 citations counted in INSPIRE as of 16 Oct 2014
	
	%\cite{Britto:2004ap}
\bibitem{Britto:2004ap} 
  R.~Britto, F.~Cachazo and B.~Feng,
  %``New recursion relations for tree amplitudes of gluons,''
  Nucl.\ Phys.\ B {\bf 715}, 499 (2005)
  [hep-th/0412308].
  %%CITATION = HEP-TH/0412308;%%
  %515 citations counted in INSPIRE as of 21 Sep 2014
%\cite{Britto:2005fq}
%\bibitem{Britto:2005fq} 
  R.~Britto, F.~Cachazo, B.~Feng and E.~Witten,
  %``Direct proof of tree-level recursion relation in Yang-Mills theory,''
  Phys.\ Rev.\ Lett.\  {\bf 94}, 181602 (2005)
  [hep-th/0501052].
  %%CITATION = HEP-TH/0501052;%%
  %608 citations counted in INSPIRE as of 23 Sep 2014

	
	%\cite{Bern:1994zx}
\bibitem{Bern:1994zx} 
  Z.~Bern, L.~J.~Dixon, D.~C.~Dunbar and D.~A.~Kosower,
  %``One loop n point gauge theory amplitudes, unitarity and collinear limits,''
  Nucl.\ Phys.\ B {\bf 425}, 217 (1994)
  [hep-ph/9403226].
  %%CITATION = HEP-PH/9403226;%%
  %808 citations counted in INSPIRE as of 21 Sep 2014
	
	%\cite{Bern:1994cg}
\bibitem{Bern:1994cg} 
  Z.~Bern, L.~J.~Dixon, D.~C.~Dunbar and D.~A.~Kosower,
  %``Fusing gauge theory tree amplitudes into loop amplitudes,''
  Nucl.\ Phys.\ B {\bf 435}, 59 (1995)
  [hep-ph/9409265].
  %%CITATION = HEP-PH/9409265;%%
  %595 citations counted in INSPIRE as of 21 Sep 2014
	
	%\cite{Neill:2013wsa}
\bibitem{Neill:2013wsa} 
  D.~Neill and I.~Z.~Rothstein,
  %``Classical Space-Times from the S Matrix,''
  Nucl.\ Phys.\ B {\bf 877}, 177 (2013)
  [arXiv:1304.7263 [hep-th]].
  %%CITATION = ARXIV:1304.7263;%%
  %4 citations counted in INSPIRE as of 21 Sep 2014
	
	%\cite{Bjerrum-Bohr:2013bxa}
\bibitem{Bjerrum-Bohr:2013bxa} 
  N.~E.~J.~Bjerrum-Bohr, J.~F.~Donoghue and P.~Vanhove,
  %``On-shell Techniques and Universal Results in Quantum Gravity,''
  JHEP {\bf 1402}, 111 (2014)
  [arXiv:1309.0804 [hep-th], arXiv:1309.0804].
  %%CITATION = ARXIV:1309.0804;%%
  %7 citations counted in INSPIRE as of 08 Nov 2014
	
	%\cite{Dixon:1996wi}
\bibitem{Dixon:1996wi} 
  L.~J.~Dixon,
  %``Calculating scattering amplitudes efficiently,''
  In *Boulder 1995, QCD and beyond* 539-582
  [hep-ph/9601359].
  %%CITATION = HEP-PH/9601359;%%
  %389 citations counted in INSPIRE as of 23 Sep 2014
	
	%\cite{Peskin:2011in}
\bibitem{Peskin:2011in} 
  M.~E.~Peskin,
  %``Simplifying Multi-Jet QCD Computation,''
  arXiv:1101.2414 [hep-ph].
  %%CITATION = ARXIV:1101.2414;%%
  %12 citations counted in INSPIRE as of 23 Sep 2014
	
	%\cite{Holstein:2008sx}
\bibitem{Holstein:2008sx} 
  B.~R.~Holstein and A.~Ross,
  %``Spin Effects in Long Range Gravitational Scattering,''
  arXiv:0802.0716 [hep-ph].
  %%CITATION = ARXIV:0802.0716;%%
  %10 citations counted in INSPIRE as of 21 Sep 2014
	%\cite{Hergt:2008jn}

\bibitem{Hergt:2008jn} 
  S.~Hergt and G.~Schaefer,
  %``Higher-order-in-spin interaction Hamiltonians for binary black holes from Poincare invariance,''
  Phys.\ Rev.\ D {\bf 78}, 124004 (2008)
  [arXiv:0809.2208 [gr-qc]].
  %%CITATION = ARXIV:0809.2208;%%
  %34 citations counted in INSPIRE as of 08 Nov 2014
	
	%\cite{Hergt:2007ha}
\bibitem{Hergt:2007ha} 
  S.~Hergt and G.~Schaefer,
  %``Higher-order-in-spin interaction Hamiltonians for binary black holes from source terms of Kerr geometry in approximate ADM coordinates,''
  Phys.\ Rev.\ D {\bf 77}, 104001 (2008)
  [arXiv:0712.1515 [gr-qc]].
  %%CITATION = ARXIV:0712.1515;%%
  %32 citations counted in INSPIRE as of 08 Nov 2014
	
	%\cite{Chakrabarti:2013xza}
\bibitem{Chakrabarti:2013xza} 
  S.~Chakrabarti, T.~Delsate and J.~Steinhoff,
  %``Effective action and linear response of compact objects in Newtonian gravity,''
  Phys.\ Rev.\ D {\bf 88}, 084038 (2013)
  [arXiv:1306.5820 [gr-qc]].
  %%CITATION = ARXIV:1306.5820;%%
  %3 citations counted in INSPIRE as of 19 Oct 2014
	
	
	%\cite{Steinhoff:2012rw}
\bibitem{Steinhoff:2012rw} 
  J.~Steinhoff and D.~Puetzfeld,
  %``Influence of internal structure on the motion of test bodies in extreme mass ratio situations,''
  Phys.\ Rev.\ D {\bf 86}, 044033 (2012)
  [arXiv:1205.3926 [gr-qc]].
  %%CITATION = ARXIV:1205.3926;%%
  %14 citations counted in INSPIRE as of 21 Sep 2014
	
	%\cite{Aguirregabiria:2001vk}
\bibitem{Aguirregabiria:2001vk} 
  J.~M.~Aguirregabiria, L.~Bel, J.~Martin, A.~Molina and E.~Ruiz,
  %``Comparing metrics at large: Harmonic versus quoharmonic coordinates,''
  Gen.\ Rel.\ Grav.\  {\bf 33}, 1809 (2001)
  [gr-qc/0104019].
  %%CITATION = GR-QC/0104019;%%
  %14 citations counted in INSPIRE as of 25 Sep 2014
	
	%\cite{Levi:2014gsa}
\bibitem{Levi:2014gsa} 
  M.~Levi and J.~Steinhoff,
  %``Leading order finite size effects with spins for inspiralling compact binaries,''
  arXiv:1410.2601 [gr-qc].
  %%CITATION = ARXIV:1410.2601;%%
	\end{thebibliography}
\end{document}